\documentclass{PoS}
\PoS{PoS(LAT2005)281}
\usepackage{epsfig}
\usepackage{rotating}

\newcommand{\Z}{{\sf Z \!\!\!\! Z}}

\newcommand{\p}{\partial}

\title{The Status of D-Theory}

\ShortTitle{The Status of D-Theory}

\author{\speaker{Uwe-Jens Wiese}
\thanks{This work is supported by the Schweizerischer Nationalfonds.}\\

Institute for Theoretical Physics, Bern University,
Sidlerstrasse 5, CH-3012 Bern, Switzerland \\
E-mail: \email{wiese@itp.unibe.ch}}

\abstract{
Field theories are usually quantized by performing a path integral over
configurations of classical fields. This is the case both in perturbation 
theory and in Wilson's nonperturbative lattice field theory. {\em D-theory} is 
an alternative nonperturbative formulation of field theory in which classical 
fields emerge from the low-energy collective dynamics of {\em discrete} quantum
variables 
(quantum spins and their gauge analogs --- quantum links) which undergo 
{\em dimensional} reduction. D-theory was developed some time ago as a discrete
approach to $U(1)$ and $SU(2)$ pure gauge theories \cite{Cha97}, extended to 
$SU(N)$ gauge theories and full QCD in \cite{Bro99,Sch01}, and also applied to
a variety of other models \cite{Bea98,Bro04}. On the practical side, 
D-theory provides a framework for the development of efficient numerical 
methods, such as cluster algorithms. For example, in the D-theory formulation 
of $CP(N-1)$ models one can simulate efficiently at non-zero chemical potential
\cite{Cha02} or at non-zero vacuum angle $\theta$ \cite{Bea04}. On the 
conceptual side, D-theory offers a natural solution for the nonperturbative 
hierarchy problem of chiral symmetry in QCD. We also take a broader 
nonperturbative view on fundamental physics and speculate that D-theory
variables --- i.e.\ quantum spins and quantum links --- may be promising 
candidates for the physical degrees of freedom that Nature has chosen to 
regularize the standard model physics at ultra-short distances.}

\FullConference{XXIIIrd International Symposium on Lattice Field Theory \\

25-30 July 2005 \\

Trinity College, Dublin, Ireland}

\begin{document}

\section{The D-Theory Formulation of $CP(N-1)$ Models}

To illustrate D-theory in a simple setting, let us consider 2-d $CP(N-1)$ 
models \cite{DAd78}. Just like 4-d QCD, these models are asymptotically free, 
they have a nonperturbatively generated massgap, as well as instantons and 
hence $\theta$-vacua. Let us imagine a toy ``world'' whose ``standard model'' 
is just a $CP(N-1)$ model with the Euclidean action
\begin{equation}
\label{CPNaction}
S[P] = \int d^2x \ \frac{1}{g^2} \mbox{Tr}[\partial_\mu P \partial_\mu P] -
i \theta Q[P].
\end{equation}
Here $P(x) \in CP(N-1) = SU(N)/U(N-1)$ is a Hermitean $N \times N$ 
matrix-valued field which obeys $P(x)^2 = P(x)$, $P(x)^\dagger = P(x)$, and 
$\mbox{Tr} P(x) = 1$. Furthermore, $g$ is the coupling constant and 
$\theta \in [- \pi,\pi]$ is the vacuum angle which multiplies the topological 
charge
\begin{equation}
\label{CPNcharge}
Q[P] = \frac{1}{2 \pi i} \int d^2x \ \epsilon_{\mu\nu} \mbox{Tr}
[P \partial_\mu P \partial_\nu P] \in \Pi_2[SU(N)/U(N-1)] = \Pi_1[U(N-1)] = 
\Pi_1[U(1)] = \Z.
\end{equation}
The model has a global $SU(N)$ symmetry $P(x)' = \Omega P(x) \Omega^\dagger$, 
with $\Omega \in SU(N)$.

Let us assume that the actual physical values of the parameters in the toy 
world are $N = 3$ and $\theta = 0$. For the fun of the argument (and happily
ignoring the antropic principle) let us further pretend that there are toy
world physicists just as puzzled about the ultimate short distance physics of 
their world as we are about our own. Our $(1+1)$-d colleagues would be quick to
figure out that their standard model is asymptotically free \cite{DAd78}, thus 
solving their ``hierarchy problem'' of why the low-energy physics takes place 
so far below the ultimate cut-off. Still, the toy world's physics community 
would remain puzzled about their ``strong CP problem'': Why is $\theta = 0$? 
The $(1+1)$-d physicists would be able to explain ``confinement'', i.e.\ the 
absence of massless excitations, as a consequence of the 
Hohenberg-Mermin-Wagner-Coleman theorem. At large $N$ they could analytically 
calculate the $\theta$-dependence and find a first order phase transition at 
$\theta = \pm \pi$ \cite{Sei84}. At finite $N \geq 3$, on the other hand, they 
would be unable to calculate the $\theta$-dependence or the massgap 
analytically. At this point, some toy world physicist may come up with the idea
to regularize the theory on the lattice. However, not unlike in our own world, 
Wilson's lattice field theory faces severe algorithmic problems. For example,
one can show that Wolff-type embedding cluster algorithms do not work 
efficiently for $CP(N-1)$ models with $N \geq 3$ \cite{Car93}. Multigrid 
methods work reasonably well, but only at $\theta = 0$ \cite{Has92}.

D-theory offers an alternative regularization that allows one to make 
substantial algorithmic progress. In the case of $CP(N-1)$ models, the discrete
D-theory variables are generalized quantum spins 
$T_x^a = \frac{1}{2} \lambda_x^a$ which generate an $SU(N)$ symmetry 
$[T_x^a,T_y^b] = i \delta_{xy} f_{abc} T_x^c$. The spins are located on the 
sites $x$ of a square lattice with spacing $a$ of size $L \times L'$, with 
$L \gg L'$ and with periodic boundary conditions. Hence, as shown in figure 1, 
we are dealing with a quantum spin ladder consisting of $n = L'/a$ transversely
coupled spin chains of length $L$. 
\begin{figure}[tb]
\vspace{0.8cm}
\epsfig{file=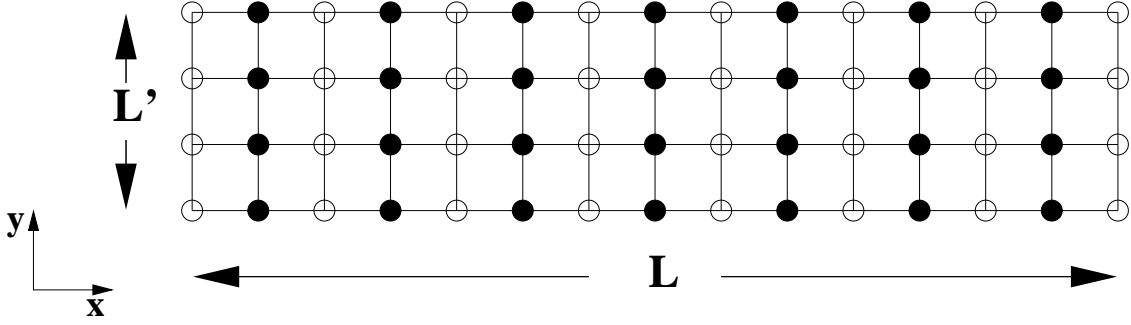,
width=2.2cm,angle=0,bbllx=0,bblly=0,bburx=225,bbury=337}
\vspace{0.2cm}
\caption{\it Spin ladder with two sublattices $A$ (open circles) and $B$ 
(filled circles).}
\end{figure}
The $x$-direction of size $L$ corresponds to the spatial dimension of the 
target $CP(N-1)$ model, while the extra $y$-dimension of finite extent $L'$ 
will ultimately disappear via dimensional reduction. We consider 
nearest-neighbor couplings which are antiferromagnetic along the chains and 
ferromagnetic between different chains. Hence, the lattice decomposes into two 
sublattices $A$ and $B$ with even and odd sites along the $x$-direction, 
respectively. The spins $T_x^a$ on sublattice $A$ transform in the fundamental 
representation $\{N\}$ of $SU(N)$, while the ones on sublattice $B$ are in the 
anti-fundamental representation $\{\overline N\}$ and are thus described by the
conjugate generators $- T_x^{a *}$. The quantum spin ladder Hamiltonian is 
given by
\begin{equation}
H = - J \sum_{x \in A} [T_x^a T_{x+\hat 1}^{a *} + T_x^a T_{x+\hat 2}^a] -
J \sum_{x \in B} [T_x^{a *} T_{x+\hat 1}^a + T_x^{a *} T_{x+\hat 2}^{a *}],
\end{equation}
where $J > 0$, and $\hat 1$ and $\hat 2$ are unit-vectors in the spatial $x$-
and $y$-directions, respectively. By construction the system has a global 
$SU(N)$ symmetry, i.e.\ $[H,T^a] = 0$, with the total spin given by 
$T^a = \sum_{x \in A} T_x^a - \sum_{x \in B} T_x^{a *}$.

One finds that, at zero temperature, the infinite system (with both 
$L, L' \rightarrow \infty$) undergoes spontaneous symmetry breaking from 
$SU(N)$ to $U(N-1)$. Hence, there are massless Goldstone bosons (spin waves)
described by fields in the coset space $SU(N)/U(N-1) = CP(N-1)$. Using chiral 
perturbation theory, the lowest-order terms in the Euclidean effective action 
for the spin waves are given by
\begin{equation}
\label{action}
S[P] = \int_0^\beta dt \int_0^L dx \int_0^{L'} dy \ \mbox{Tr} 
\{\rho_s' \p_y P \p_y P +
\rho_s [\p_x P \p_x P + \frac{1}{c^2} \p_t P \p_t P] -
\frac{1}{a} P \p_x P \p_t P\}.
\end{equation}
Here $\beta = 1/T$ is the inverse temperature, $\rho_s$ and $\rho_s'$ are spin 
stiffness parameters for the $x$- and $y$-direction, respectively, and $c$ is 
the spin wave velocity. The last term in the integrand of eq.(\ref{action}) is 
purely imaginary and is related to the topological charge $Q[P]$, which is a
$y$-independent integer. Hence, the $y$-integration in the last term of 
eq.(\ref{action}) can be performed trivially. This yields $i \theta Q[P]$ where
the vacuum angle is given by $\theta = L' \pi/a = n \pi$. Here $a$ is the 
lattice spacing of the quantum spin ladder and $L'/a = n$ is the number of 
transversely coupled spin chains. Hence, for even $n$ the vacuum angle is 
trivial, and for odd $n$ it corresponds to $\theta = \pi$.

While the infinite $(2+1)$-d system has massless Goldstone bosons, the
Coleman-Hohenberg-Mermin-Wagner theorem forbids the existence of massless
excitations once the $y$-direction is compactified to a finite extent $L'$.
As a consequence, the Goldstone bosons then pick up a nonperturbatively
generated massgap $m = 1/\xi$ and thus have a finite correlation length $\xi$.
Interestingly, for sufficiently many transversely coupled chains, the 
correlation length becomes exponentially large 
$\xi \propto \exp(4 \pi L' \rho_s/c N) \gg L'$, and the system undergoes 
dimensional reduction to the $(1+1)$-d $CP(N-1)$ field theory with the action
\begin{equation}
S[P] = \int_0^\beta dt \int_0^L dx \ 
\mbox{Tr}\{\frac{1}{g^2} [\p_x P \p_x P + \frac{1}{c^2} \p_t P \p_t P] - 
n P \p_x P \p_t P\}.
\end{equation}
The coupling constant of the dimensionally reduced theory is given by $1/g^2 =
L'\rho_s/c$. This type of dimensional reduction is well-known for
antiferromagnets \cite{Cha88,Has91}.

When regularized using D-theory a highly efficient loop-cluster algorithm can 
be applied to $CP(N-1)$ models \cite{Bea04}. In this way large correlation
lengths of up to 250 lattice spacings have been simulated with no indication of
critical slowing down. It has also been possible to simulate at non-zero 
chemical potential \cite{Cha02} and at vacuum angle $\theta = \pi$ 
\cite{Bea04}. In this way, it was shown for several $N \geq 3$ that there is a 
first order phase transition at $\theta = \pi$ at which charge conjugation gets
spontaneously broken. Algorithmic developments for the D-theory formulation of 
QCD are currently under intensive investigation. Ironically, while the
meron-cluster algorithm provides a very efficient method to simulate dynamical 
fermions \cite{Cha99}, at present the simulation of quantum links still causes 
severe problems. 

When the experimentalists in our toy world will be able to probe the shortest
distance scales, they may discover the (perhaps somewhat disappointing) fact
that they actually live on, let us say, $n = 10$ transversely coupled chains of
$SU(3)$ spins. This also solves their ``strong CP-problem'': $\theta = 0$ 
because $n$ is even. In the following, we like to speculate that our own world 
may in some respects be not so different from the toy example.

\section{A Nonperturbative View on Fundamental Physics}

Dimensional regularization provides an elegant but unphysical regularization,
which is very useful in QCD, but it defines the theory only in perturbation 
theory. In a chiral gauge theory like the full standard model, dimensional 
regularization of $\gamma_5$ is subtle beyond one loop. Such subtleties provide
a first perturbative glance at a deep problem that becomes apparent when one 
regularizes theories with a chiral symmetry beyond perturbation theory. In 
Wilson's lattice field theory, due to fermion doubling, chiral symmetry has 
posed severe problems for many years. In particular, using Wilson fermions, 
i.e.\ removing the doubler fermions by breaking chiral symmetry explicitly, 
causes a severe nonperturbative hierarchy problem for fermions \cite{Cha04}. 
Without unnatural fine-tuning of the bare fermion mass it is then impossible to
obtain light fermions. This problem has sometimes been viewed as a deficiency 
of the lattice regularization. In particular, the global chiral symmetry of 
massless QCD is usually taken for granted because it can easily be maintained 
in continuum regularization schemes. A continuum field theorist could 
``explain'' the presence of (almost) massless fermions in Nature by the 
existence of (an approximate) chiral symmetry which protects the quark masses 
from running to the cut-off scale. We like to stress that this perturbative 
point of view of the problem is rather limited, and may even prevent us from 
drawing some far-reaching conclusions about the physics at ultra-short distance
scales. 

Remarkably, the hierarchy problem of the nonperturbative regularization of 
chiral symmetry has found an elegant solution in terms of Kaplan's domain wall 
fermions \cite{Kap92,Sha93}. Massless 4-d fermions then arise naturally as 
states localized on a domain wall embedded in a 5-d space-time, while fermion 
doublers are still removed by a 5-d Wilson term. Narayanan's and Neuberger's 
closely related overlap fermions \cite{Nar93} are also deeply related to the 
physics of an extra dimension. Hence, solving the nonperturbative hierarchy 
problem of fermions may require at least one extra dimension. Without invoking 
extra dimensions, at a nonperturbative level we can presently not understand 
how fermions can be naturally light. The existence of light fermions in Nature 
may thus be a concrete hint to the physical reality of extra dimensions. In 
particular, we don't need string theory or other physics beyond the standard 
model to motivate extra dimensions. The mere existence of light fermions in 
Nature is evidence already. This important hint from nonperturbative physics is
indeed easily missed when one considers chiral symmetry only in a perturbative 
context.

Why is the weak scale so much smaller than the GUT or Planck scale? This is the
gauge hierarchy problem of the standard model. In contrast to the 
nonperturbative hierarchy problem of chiral symmetry, the gauge hierarchy 
problem manifests itself already in perturbation theory and is therefore widely
appreciated. The presently most popular potential solution of this problem 
relies on supersymmetry. However, from a nonperturbative point of view this 
``solution'' is not yet satisfactory. Beyond perturbation theory, namely on the
lattice, a priori supersymmetry is as undefined as chiral symmetry was before 
Kaplan constructed lattice domain wall fermions. Lattice scalar field theory 
suffers from the same hierarchy problem as the continuum theory, i.e.\ without 
unnatural fine-tuning of the bare mass, the vacuum value of the Higgs field 
remains at the lattice cut-off. Obtaining supersymmetry in the continuum limit 
thus requires fine-tuning of the bare scalar mass. Of course, as long as the 
nonperturbative construction of supersymmetry itself requires unnatural 
fine-tuning, it cannot solve the hierarchy problem. In the worst case, the 
supersymmetric extension of the standard model may just be a perturbative 
illusion which does not arise naturally, i.e.\ without fine-tuning, in a 
nonperturbative context. Until now, other than for chiral symmetry, Nature has 
not yet provided us with experimental evidence for supersymmetry. While this 
may well change in the near future, one can presently not be sure that 
supersymmetric extensions of the standard model even exist naturally beyond 
perturbation theory.

Ultimately, the divergences of quantum field theory imply that the concept of a
classical field (originally developed for classical electrodynamics) breaks 
down at ultra-short distances. In particular, Dirac continued to point out that
he was unsatisfied with the formal procedures of removing singularities in the 
perturbative treatment of QED \cite{Dir83}. Indeed, it is hard to imagine that 
classical fields are the truly fundamental physical degrees of freedom that 
Nature has chosen to regularize particle physics. No matter if there are 
strings, branes, or some tiny wheels turning around at the Planck scale, Nature
must have found a concrete way to regularize gravity as well as the standard 
model physics at ultra-short distances. Of course, the identification of the 
ultimate hardware on which the basic laws of Nature are implemented is a very 
difficult task which may or may not be within reach of physics in the 
foreseeable future. Here we like to speculate that discrete variables, namely 
quantum spins and their gauge analogs --- quantum links --- may be promising 
candidates for Nature's most fundamental degrees of freedom. 

D-theory provides a framework in which the familiar classical fields emerge 
naturally from discrete quantum variables that undergo dimensional reduction.
In the D-theory formulation of QCD \cite{Bro99} a fifth dimension is not only 
needed to obtain naturally light quarks, but also to assemble 4-d gluons out of
5-d quantum links. If we take the existence of light fermions as a hint to the
reality of extra dimensions, we should take these dimensions seriously also for
the gauge fields. Just like Wilson's parallel transporters, quantum links are 
$N \times N$ matrices which transform appropriately under $SU(N)$ gauge 
transformations. However, like the components of a quantum spin, their 
matrix elements are operators (in the QCD case generators of $SU(2N)$). The 
collective dynamics of the discrete quantum link variables may give rise to a
5-d non-Abelian Coulomb phase which is analogous to the 3-d phase with massless
Goldstone bosons. Just as Goldstone bosons pick up a mass as a consequence of
the Hohenberg-Mermin-Wagner-Coleman theorem when the third direction is
compactified, with a compact fifth dimension Coulombic gluons form glueballs 
and are thus confined \cite{Cha97}.

Until now D-theory has not been widely recognized as a potential framework for 
a truly fundamental theory. Although this is highly speculative, we like to 
point out that D-theory indeed offers room for nonperturbative thought on 
fundamental physics alternative to string theory. Of course, the present
constructions with a rigid lattice and just one extra dimension may not be 
sufficient, but the idea that the most fundamental degrees of freedom are 
discrete quantum variables may lead to fruitful developments. Regularizing the 
full standard model or gravity in the D-theory framework represent great 
challenges that seem worth facing.


\begin{thebibliography}{99}

\bibitem{Cha97}
S.~Chandrasekharan and U.-J.~Wiese, Nucl.\ Phys.\ B492 (1997) 455
[arXiv:hep-lat/9609042].

\bibitem{Bro99}
R.~Brower, S.~Chandrasekharan, and U.-J.~Wiese, Phys.\ Rev.\ D60 (1999) 094502
[arXiv:hep-lat/9704106].

\bibitem{Sch01}
B.~Schlittgen and U.-J.~Wiese, Phys.\ Rev.\ D63 (2001) 085007 
[arXiv:hep-lat/0012014].

\bibitem{Bea98}
B.~B.~Beard, R.~J.~Birgeneau, M.~Greven, and U.-J.~Wiese, Phys.\ Rev.\ Lett.\ 
80 (1998) 1742 [arXiv:cond-mat/9709110].

\bibitem{Bro04}
R.~C.~Brower, S.~Chandrasekharan, S.~Riederer, and U.-J.~Wiese,  Nucl.\ Phys.\ 
B693 (2004) 149 [arXiv:hep-lat/0309182].

\bibitem{Cha02}
S.~Chandrasekharan, B.~Scarlet, and U.-J.~Wiese, Comput.\ Phys.\ Commun.\ 147
(2002) 388 [arXiv:hep-lat/0110215].

\bibitem{Bea04}
B.~B.~Beard, M.~Pepe, S.~Riederer, and U.-J.~Wiese, Phys.\ Rev.\ Lett.\ 94
(2005) 010603 [arXiv:0406040].

\bibitem{DAd78}
A.~D'Adda, P.~Di~Vecchia, and M.~L\"uscher, Nucl.\ Phys.\ B146 (1978) 63;
Nucl.\ Phys.\ B152 (1979) 125.

\bibitem{Sei84}
N.~Seiberg, Phys.\ Rev.\ Lett.\ 53 (1984) 637.

\bibitem{Car93}
S.~Caracciolo et al., Nucl.\ Phys.\ B403 (1993) 475 [arXiv:hep-lat/9205005].

\bibitem{Has92}
M.~Hasenbusch and S.~Meyer, Phys.\ Rev.\ Lett.\ 68 (1992) 435, 
Phys.\ Rev.\ D45 (1992) 4376.

\bibitem{Cha88}
S.~Chakravarty, B.~I.~Halperin, and D.~R.~Nelson, Phys.\ Rev.\ Lett. 60
(1988) 1057.

\bibitem{Has91}
P.~Hasenfratz and F.~Niedermayer, Phys.\ Lett.\ B268 (1991) 231.

\bibitem{Cha99}
S.~Chandrasekharan and U.-J.~Wiese, Phys.\ Rev.\ Lett.\ 83 (1999) 3116
[arXiv:hep-lat/9902128].

\bibitem{Cha04}
S.~Chandrasekharan and U.-J.~Wiese, Prog.\ Part.\ Nucl.\ Phys.\ 53 (2004) 373
[arXiv:hep-lat/0405024].

\bibitem{Kap92}
D.~B.~Kaplan, Phys.\ Lett.\ B288 (1992) 342 [arXiv:hep-lat/9206013].

\bibitem{Sha93}
Y.~Shamir, Nucl.\ Phys.\ B406 (1993) 90 [arXiv:hep-lat/9303005].

\bibitem{Nar93}
R.~Narayanan and H.~Neuberger, Phys.\ Lett.\ B302 (1993) 62 
[arXiv:hep-lat/9212019].

\bibitem{Dir83}
P.~A.~M.~Dirac, ``The Inadequacies of Quantum Field Theory'', printed in
``Paul Adrien Maurice Dirac: Reminiscences about a great Physicist'', eds.\
B.~N.~Kursunoglu and E.~P.~Wigner, Cambridge University Press (1987).

\end{thebibliography}
\end{document}